\documentclass[twocolumn,showpacs,floatfix]{revtex4}%
\usepackage{graphicx}%
\usepackage{amsmath}%
\usepackage{amsfonts}%
\usepackage{amssymb}
\usepackage{bm}

\def\s{{\sigma}}
\def\e{{\epsilon}}
\def\k{{ {\bm k} }}

\def\Q{{ {\bm Q} }}

\def\w{{\omega}}
\def\a{{\alpha}}
\def\b{{\beta}}

\allowdisplaybreaks[4]

\begin{document}
\title{Nodal gap structure in Fe-based superconductors \\
due to the competition between orbital and spin fluctuations.
}
\author{Tetsuro \textsc{Saito}$^{1}$,
Seiichiro \textsc{Onari}$^{2}$,
and Hiroshi \textsc{Kontani}$^{1}$
}

\date{\today }

\begin{abstract}
To understand the origin of the nodal gap structure realized in 
BaFe$_2$(As,P)$_2$,
we study the three-dimensional gap structure based on the three-dimensional 
ten-orbital Hubbard model with quadrupole interaction.
In this model, strong spin and orbital fluctuations develop
by using the random-phase-approximation.
By solving the Eliashberg gap equation,
we obtain the fully-gapped $s$-wave state with (without) sign reversal
between hole-like and electron-like Fermi surfaces
due to strong spin (orbital) fluctuations, so called the $s_\pm$-wave
($s_{++}$-wave) state.
When both spin and orbital fluctuations strongly develop,
which will be realized near the orthorhombic phase, 
we obtain the nodal $s$-wave state in the crossover region
between $s_{++}$-wave and $s_\pm$-wave states.
The obtained nodal $s$-wave state possesses
the loop-shape nodes on electron-like Fermi surfaces,
due to the competition between attractive and repulsive interactions
in $\k$-space.
In contrast, the SC gaps on the hole-like Fermi surfaces are fully-gapped
due to orbital fluctuations.
The present study explains the main characters of the 
anisotropic gap structure in BaFe$_2$(As,P)$_2$ observed experimentally.
\end{abstract}

\address{
$^1$ Department of Physics, Nagoya University,
Furo-cho, Nagoya 464-8602, Japan. 
\\
$^2$ Department of Applied Physics, Nagoya University,
Furo-cho, Nagoya 464-8602, Japan. 
}
 

\sloppy

\maketitle

\section{Introduction}

\begin{figure}[!htb]
\includegraphics[width=0.6\linewidth]{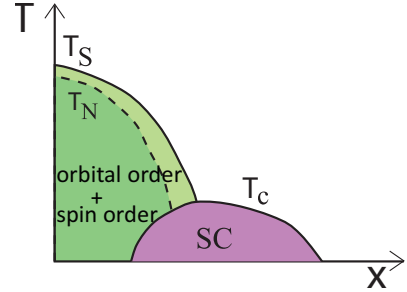}
\caption{
(Color online)
A typical phase diagram for Fe-based superconductors.
$T_S$ is the structure transition temperature,
which is expected to be induced by orbital polarization
$n_{xz}>n_{yz}$ according to the ARPES measurements
and the sizable softening of $C_{66}$.
$T_N$ is the magnetic transition temperature.
}
\label{fig:phasediagram}
\end{figure}

Since the discovery of Fe-based high-$T_{\rm c}$ superconductors
by Kamihara {\it et al}.\cite{Hosono},
their many-body electronic properties have been studied very intensively.
Figure \ref{fig:phasediagram} shows a typical phase diagram of 
Fe-based superconductors.
In the under-doped regime, 
the second-order orthorhombic (O) structure transition occurs at $T_{\rm S}$,
and the stripe-type magnetic order is realized at $T_{\rm N}\lesssim T_{\rm S}$.
In the O phase, the orbital polarization $n_{xz}\ne n_{yz}$ is realized,
where $n_{xz(yz)}$ is the filling of $d_{xz(yz)}$ orbital
\cite{ARPES-Shen}.
Also, sizable softening of shear modulus $C_{66}$ 
\cite{Fernandes,Yoshizawa,Goto}
and the renormalization of phonon velocity
\cite{neutron-phonon}
indicate the development of orbital fluctuations near the 
orthorhombic phase.
Strong spin fluctuations are also observed near the 
magnetic ordered phase.

These main characters of the phase diagram should be understood
in order to clarify the mechanism of superconductivity.
Theoretical studies of orbital polarization had been proposed
in Refs. \cite{OO1,OO2,OO3}.
However, {\it non-magnetic} structure transition cannot be explained 
based on the Hubbard model, once we apply the mean-field approximation 
or random-phase-approximation (RPA).
To solve this problem, we had recently improved the RPA
by including the vertex correction (VC) for the susceptibility
that is dropped in the RPA \cite{Onari-SCVC}:
By applying this self-consistent VC (SC-VC) method
to the Hubbard model for Fe-based superconductors,
both spin and orbital fluctuations mutually develop, 
and both the O structure transition and the softening of $C_{66}$
can be explained.
Note that the ``electronic nematic state'' 
with large in-plane anisotropy of resistivity or magnetization
well above $T_S$ \cite{Fisher,Matsuda,Eisaki}
also indicates the occurrence of the (local) 
orbital order \cite{Inoue-Nematic}.

The phase diagram in Fig. \ref{fig:phasediagram}
indicates that both spin and orbital fluctuations could be
closely related to the mechanism of high-$T_{\rm c}$.
Up to now, the spin-fluctuation-mediated $s_\pm$-wave state
\cite{Kuroki,Mazin,Hirschfeld,Chubukov}
and orbital-fluctuation-mediated $s_{++}$-wave state
\cite{Kontani-RPA,Saito-RPA}
had been studied based on the multiorbital models.
The $s_\pm$-wave state has the 
sign reversal of the gap between hole-like Fermi surfaces (h-FSs) 
and electron-like Fermi surfaces (e-FSs),
whereas the $s_{++}$-wave state has no sign reversal.
Experimentally, the robustness of 
$T_{\rm c}$ against impurities in many Fe-based superconductors
\cite{Sato-imp,Nakajima,Li,Paglione}
indicates the realization of the  $s_{++}$-wave state, at least in
``dirty'' compounds with high residual resistivity \cite{Onari-imp}.
Also, the ``resonance-like'' hump structure 
in the neutron inelastic scattering \cite{Onari-resonance}
can be explained by considering the energy dependence of the 
inelastic scattering if the $s_{++}$-wave state is realized.

Although fully-gapped $s$-wave state is realized in many 
optimally-doped high-$T_{\rm c}$ compounds, nodal $s$-wave state 
(accidental node is not protected by symmetry) is also observed 
in some compounds with lower $T_{\rm c}$ 
\cite{kappa-node}.
The appearance of the accidental node strongly
indicates the presence of ``competing pairing interactions''
\cite{Kontani-B}.
In the spin fluctuation scenario,
if the $xy$-orbital hole-pocket is under the Fermi level,
spin fluctuations in $xz+yz$-orbitals develop at $\Q=(\pi,0)$
while those in $xy$-orbital develop at $\Q'=(\pi,\pi/2)$,
and this frustration gives a nodal gap structure
around the $xy$-orbital part on the e-FS
\cite{Scalapino}.
This mechanism was studied in detail in Ref. \cite{Maiti-nodal}
by introducing phenomenological pairing interaction.
However, the $xy$-orbital h-FS presents in real compounds.
In this case, fully-gapped $s_\pm$-wave state is obtained by the RPA
since spin fluctuations develop at $\Q=(\pi,0)$ in all $d$-orbitals,
consistently with neutron experiments
\cite{Kuroki-PRB}.

Very interestingly, in optimally-doped BaFe$_2$(As,P)$_2$,
nodal gap structure with high-$T_{\rm c}$ ($\sim30$K) is realized.
The superconducting (SC) gaps on the three h-FSs are fully-gapped 
and almost orbital-independent both in the $k_z=\pi$ plane
\cite{Shimo-Science} and in the $k_z=0$ plane \cite{Yoshida},
consistently with the orbital fluctuation scenario in Ref. \cite{Saito-RPA}.
Also, loop-shape nodes on the e-FSs are observed by angle-resolved 
thermal conductivity measurement in the vortex state \cite{Yamashita} 
and ARPES measurements \cite{Shimo-Science,Yoshida}.
These results indicate the existence of competing 
pairing interactions, and the study of these facts
would be significant to understand the mechanism of 
high-$T_{\rm c}$ superconductivity.

On the other hand,  the ARPES measurement by Ref. \cite{Feng} 
reported the horizontal node on the $z^2$-orbital e-FS around the Z point
in BaFe$_2$(As,P)$_2$, contrary to the reports by 
Refs. \cite{Shimo-Science,Yoshida}.
This result is consistent with the prediction of the theory of the
spin-fluctuation-mediated $s_\pm$-wave state in Ref. \cite{Suzuki}.
However, the existence of the horizontal node would be inconsistent
with the large in-plane field angle dependence of the thermal conductivity
reported in Ref. \cite{Yamashita}.
Also, very small $T$-linear term in the specific heat in the SC state
would not be compatible to the presence of nodes 
on heavy hole-like FSs \cite{Kim,Wang}.

In this paper, we theoretically study the origin of the nodal gap structure
in BaFe$_2$(As,P)$_2$, in order to obtain a significant information 
of the pairing mechanism of Fe-based superconductors.
For this purpose, we construct the three-dimensional (3D) ten-orbital 
tight-binding model for BaFe$_2$(As,P)$_2$, 
and calculate the dynamical spin and orbital susceptibilities 
due to the combination of Coulomb and quadrupole interactions.
By solving the Eliashberg gap equation,
we obtain the fully-gapped $s_{\pm}$-wave ($s_{++}$-wave) state 
due to strong spin (orbital) fluctuations.
When both spin and orbital fluctuations strongly develop, 
which will be realized near the O phase,
nodal $s$-wave state with loop-shape nodes on the e-FSs is realized
due to the competition between attractive and repulsive interactions.
It is realized 
during a smooth crossover between $s_{++}$- and $s_\pm$-wave states
\cite{Kontani-RPA,Hirsch-cross}.
In contrast, the SC gaps on the h-FSs are fully-gapped due to orbital fluctuations.
Thus, the present study explains the main characters of the 
gap structure in BaFe$_2$(As,P)$_2$.

In Refs. \cite{Kontani-RPA,Saito-RPA,Saito-RPA2,Kontani-AL,Kontani-SSC}
present authors have shown that small quadrupole interaction 
induced by Fe-ion oscillations
gives rise to the large antiferro- and ferro-orbital fluctuations.
In addition, we had developed the {\it spin+orbital} 
fluctuation theory in multiorbital Hubbard model 
by including the VCs to the susceptibilities,
which are neglected in the RPA \cite{Onari-SCVC}.
It was found that the Aslamazov-Larkin type VC
due to Coulomb interaction produces large effective 
quadrupole interaction.
The emergence of the orbital fluctuations due to the VC
is also recognized in a simple two-orbital model,
using the self-consistent VC method \cite{Ohno-SCVC} 
as well as newly developed two-dimensional 
renormalization group method (RG+cRPA method) \cite{Tsuchiizu}.

\begin{figure}[!htb]
\includegraphics[width=0.8\linewidth]{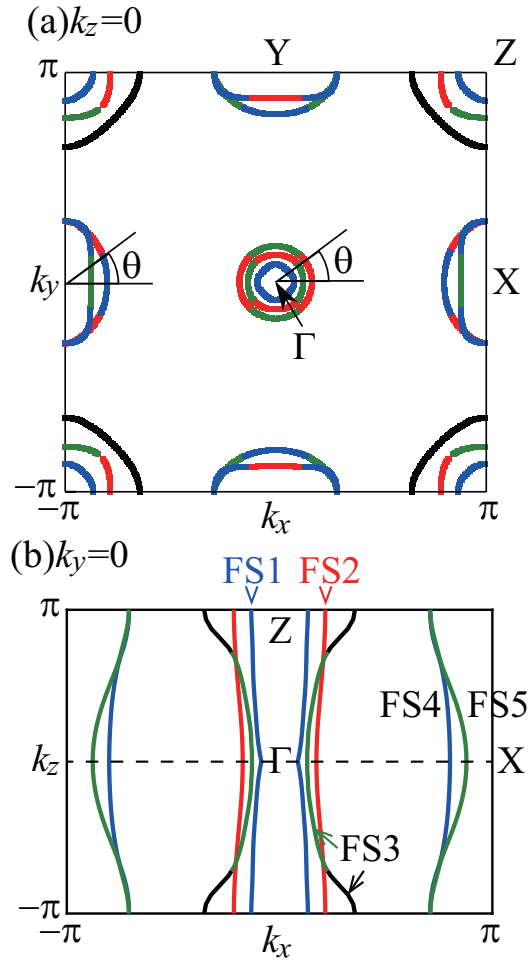}
\caption{
The Fermi surfaces in the (a)$k_z=0$ plane and (b)$k_y=0$ plane
of the present ten orbital model for the filling $n=6.0$.
The green, red, blue and black lines correspond to
$xz$, $yz$, $xy$ and $z^2$ orbitals, respectively.
In (b), there are three h-FSs (FS1, FS2 and FS3) and four e-FSs (FS4 and FS5).
}
\label{fig:Fermi}
\end{figure}

In Sec. \ref{sec:formulatoin},
we introduce the three-dimensional ten-orbital 
tight-binding model, which contains two Fe-sites in each unit cell.
We analyze this model based on the RPA, by taking both the
Coulomb and quadrupole interactions into account.
The latter interaction originates from the Coulomb interaction
beyond the RPA, described by the vertex corrections.
In Sec. \ref{sec:SC}, we analyze the SC gap equation
for various model parameters, and derive the loop-shape nodes
on e-FSs due to the competition between orbital and spin fluctuations.
Some discussions and the summary are presented in Secs. \ref{sec:discussion}
and \ref{sec:summary}, respectively.
In Appendix A, we show the obtained orbital fluctuations
that give the $s_{++}$-wave state with nearly isotropic gap functions 
on the three h-FSs.
In Appendix B, we discuss the SC state in heavily electron-doped systems
when the $xy$-orbital h-FS disappears.

\section{Formulation}
\label{sec:formulatoin}

In this paper, we set $x$ and $y$ axes parallel to the nearest Fe-Fe bonds,
and the orbital $z^2$, $xz$, $yz$, $xy$, and $x^2-y^2$ are denoted as
1, 2, 3, 4, and 5 respectively.
First, we perform the local-density-approximation (LDA) band calculation
for BaFe$_2$As$_2$ and BaFe$_2$P$_2$ using WIEN2K code based on the experienced crystal structure.
Next, we derive the ten-orbital tight-binding model that reproduces the LDA band structure
and its orbital character using WANNIER90 code and WIEN2WANNIER interface. \cite{win2wannier}
Using the obtained two sets of tight-binding parameters 
(hopping integrals and on-site energies),
the parameters of BaFe$_2$(As$_{1-x}$P$_x$)$_2$ are well approximated by
making a linear combination of them with a ratio of $1-x : x$
\cite{Suzuki}.
In this paper, we use the tight-binding parameters for $x=0.30$.
The obtained kinetic term is given as
\begin{multline}
\hat{H}^{0}= \sum_{ab \alpha \beta l m \sigma}
t^{a \alpha, b \beta}_{lm} c^{\dagger}_{la \alpha , \sigma} c_{m b \beta , \sigma} \\
= \sum_{ab \alpha \beta l m \sigma} \sum_{\bm{k}} t^{a \alpha, b \beta}_{lm} e^{i \bm{k} \cdot (\bm{R}_{a, \alpha} - \bm{R}_{b, \beta})}
c^{\dagger}_{l \alpha , \sigma}(\bm{k}) c_{m \beta , \sigma} (\bm{k})
\end{multline}
where $a,b$ represent the unit cell, $\alpha , \beta$ ($=A$, $B$) represent the two Fe sites,
$l,m = 1-5$ represent the $d$ orbital, and $\sigma = \pm 1$ is the spin index.
$\bm{R}_{a, \alpha}$ is the position of Fe-site,
$c^{\dagger}_{la \alpha ,\sigma}$ is the creation operator of the $d$ electron,
and $t^{a \alpha, b \beta}_{lm}$ with $a=b$ and $\alpha = \beta$ ($a \ne b$ or $\alpha \ne \beta$)
is the local potential (hopping integral).

However, the $xy$-orbital h-FS given by the LDA
is too small compared to the experimental results by ARPES measurements.
In order to increase the size of $xy$-orbital h-FS,
we introduce the following orbital-dependent potential term
around the $\Gamma$-point:
\begin{multline}
\hat{H}^{\mathrm{kin}}
= \hat{H}^{0} \\
+ \sum_{l \alpha , \sigma} \sum_{\bm{k}} e_l \left[ \frac{\cos{k_x} \cos{k_y} +1}{2} \right] c^{\dagger}_{l \alpha , \sigma}(\bm{k}) c_{l \alpha , \sigma} (\bm{k}),
\end{multline}
where $e_l$ is the energy shift of the orbital-$l$ at $\Gamma$-point.
We put $e_{xy}=0.02$ eV, $e_{xz}=e_{yz}=-0.01$ eV and the others are 0.
The FSs in this model are composed of 
three h-FSs around $\Gamma$-point and four e-FSs
around $X$- and $Y$-points.
Figure \ref{fig:Fermi} show the obtained FSs in the
(a) $k_z=0$ and (b) $k_y=0$ planes, respectively.
The electron filling per Fe-site is $n=6.0$.
In Fig. \ref{fig:Fermi} (b), there are three h-FSs (FS1, FS2 and FS3) 
and two e-FSs (FS4 and FS5).
We call FS4 (FS5) the outer (inner) e-FS.

Next, we explain the interaction term.
We introduce both the Coulomb interaction ($U$, $U'$, $J=(U-U')/2$) 
and quadrupole interaction.
The latter are induced by the electron-phonon (e-ph) interaction
due to Fe ion oscillations as follows,\cite{Kontani-AL}
\begin{equation}
 \begin{split}
V_{\rm{quad}}=- & g_1(\w_l) \sum_i^{\mathrm{site}} \left( 
{\hat O}^i_{yz} \cdot{\hat O}^i_{yz} + {\hat O}^i_{xz} \cdot{\hat O}^i_{xz} \right) \\
- & g_2(\w_l) \sum_{i}^{\mathrm{site}} \left( {\hat O}^i_{xy} \cdot{\hat O}^i_{xy} \right),
 \end{split}
 \label{eqn:Hint}
\end{equation}
where $g_i(\w_l)=g_i \cdot \w_{\rm D}^2/(\w_l^2+\w_{\rm D}^2)$, and 
$g_i=g_i(0)$ is the quadrupole interaction at $\omega_n=0$.
$\omega_{\mathrm{D}}$ is the cutoff energy of the quadrupole interaction.
$\hat{O}_{\Gamma}$ is the quadrupole operator introduced 
in Ref. \cite{Kontani-RPA}, which will be shown in Appendix A.
$\hat{V}_{\mathrm{quad}}$ has many non-zero off-diagonal elements as explained in 
Refs.\cite{Kontani-RPA}.
As explained in Ref. \cite{Kontani-RPA},
$g_1$ ($g_2$) is induced by in-plane (out-of-plane) Fe-ion oscillations.
In this paper, we put $g_1 = g_2 =g$ unless otherwise noted.
Also, the Aslamazov-Larkin type VC
due to Coulomb interaction produces large effective 
quadrupole interaction $g_1$ \cite{Onari-SCVC}.
Thus, the quadrupole interaction in eq. (\ref{eqn:Hint}) 
is derived from both the VC and e-ph interaction.

Now, we perform the RPA for the present model,
by using $32 \times 32 \times 16 \ \bm{k}$ meshes.
The irreducible susceptibility in the ten orbital model is given by
\begin{equation} 
\chi^{(0) \alpha \beta}_{ll',mm'} \left( q \right) =
- \frac{T}{N} \sum_k G_{lm}^{\alpha \beta} \left( k+ q \right) G_{m' l'}^{\beta \alpha} \left( k \right),
\end{equation}
where $q = ( \bm{q}, \omega_l )$ and $k=( \bm{k} , \epsilon_n)$.
$\epsilon_n = (2n + 1) \pi T$ and $\w_l = 2l \pi T$ 
are the fermion and boson Matsubara frequencies.
$\hat{G} ( k ) = [ i \epsilon_n + \mu - \hat{h}^{\mathrm{kin}}_{\bm{k}} ]^{-1}$
is the \textit{d} electron Green function in the orbital basis,
where $\hat{h}^{\mathrm{kin}}_{\bm{k}}$ is the matrix elements of 
$\hat{H}^{\mathrm{kin}}$ and $\mu$ is the chemical potential.
Then, the susceptibilities for spin and charge sectors in the RPA are given by 
\cite{Takimoto}
\begin{gather} 
\hat{\chi}^{\mathrm{s}} \left( q \right) = \frac{\hat{\chi}^{(0)} \left( q \right)}{\hat{1} - \hat{\Gamma}^{\mathrm{s}} \hat{\chi}^{(0)} \left( q \right)}, 
\label{eqn:chis} \\
\hat{\chi}^{\mathrm{c}} \left( q \right) = \frac{\hat{\chi}^{(0)} \left( q \right)}{\hat{1} - \hat{\Gamma}^{\mathrm{c}} (\omega_l) \hat{\chi}^{(0)} \left( q \right)},
\label{eqn:chic}
\end{gather}
where
\begin{equation}
(\Gamma^{\mathrm{s}})^{\alpha \beta}_{l_{1}l_{2},l_{3}l_{4}} = \delta_{\alpha,\beta} \times \begin{cases}
U, & l_1=l_2=l_3=l_4 \\
U' , & l_1=l_3 \neq l_2=l_4 \\
J, & l_1=l_2 \neq l_3=l_4 \\
J' , & l_1=l_4 \neq l_2=l_3 \\
0 , & \mathrm{otherwise}
\end{cases}
\end{equation}
\begin{equation}
\hat{\Gamma}^{\mathrm{c}} ( \omega_l )= -\hat{C} - 2\hat{V}_{\mathrm{quad}}( \w_l ),
\label{eqn:Gc}
\end{equation}
\begin{equation}
(C)^{\alpha \beta}_{l_{1}l_{2},l_{3}l_{4}} = \delta_{\alpha,\beta} \times \begin{cases}
U, & l_1=l_2=l_3=l_4 \\
-U'+2J , & l_1=l_3 \neq l_2=l_4 \\
2U' - J , & l_1=l_2 \neq l_3=l_4 \\
J' , &l_1=l_4 \neq l_2=l_3 \\
0 . & \mathrm{otherwise}
\end{cases}
\end{equation}
where $\a,\b=A,B$.

In the RPA, the enhancement of the spin susceptibility $\hat{\chi}^{\mathrm{s}}$
is mainly caused by the intra-orbital Coulomb interaction $U$,
using the ``intra-orbital nesting'' of the FSs.
On the other hand, the enhancement of $\hat{\chi}^{\mathrm{c}}$ in the present model
is caused by the quadrupole-quadrupole interaction in eq. (\ref{eqn:Hint}),
utilizing the ``inter-orbital nesting'' of the FSs.
The magnetic (orbital) order is realized 
when the spin (charge) Stoner factor $\alpha_{\mathrm{s}(\mathrm{c})}$,
which is the maximum eigenvalue of $\hat{\Gamma}^{\mathrm{s}(\mathrm{c})} \hat{\chi}^{(0)} ( \bm{q} , 0)$, is unity.
When $n=6.0$, the critical value of $U$ is $U_{\mathrm{cr}}=1.18$ eV,
and the critical value of $g$ is $g_{\mathrm{cr}}= 0.23$ eV for $U=0$.
Hereafter, we set the unit of energy as eV.

Next, we explain the linearized Eliashberg equation.
In order to obtain the fine momentum dependence of the SC gap,
we concentrate on the gap functions only on the FSs
as done in Ref. \cite{Scalapino}:
We used $40 \times 16 \ \bm{k}$ points for each Fermi surface sheet.
In the presence of dilute impurities ($n_{\mathrm{imp}} \ll 1$),
the linearized Eliashberg equation is given as \cite{Scalapino}
\begin{multline}
Z^{i} (\bm{k}, \e_n ) \lambda_{\mathrm{E}} \Delta^{i} ( \bm{k} , \e_n ) \\
= \frac{\pi T}{(2 \pi )^3} \sum_{\e_m} \sum_{j}^{\mathrm{FS}} \int_{\mathrm{FS}j} \frac{d \bm{k}'_{\mathrm{FS}j}}{v^{j} ( \bm{k}')}
V^{ij}( \bm{k} , \bm{k}' , \e_n - \e_m ) \\
\times \frac{\Delta^{j}( \bm{k}' , \e_m )}{| \e_m |}
+ \delta \Sigma^i_{\mathrm{a}} ( \bm{k} , \e_n ),
\label{eqn:Eliasheq}
\end{multline}
where $\lambda_{\mathrm{E}}$ is the eigenvalue that reaches unity at $T=T_{\mathrm{c}}$.
$i$ and $j$ denote the FSs, and 
$ \Delta^{i} ( \bm{k} , \e_n )$ is the 
gap function on the $i$-th FS (FS$i$) at the Fermi momentum $\bm{k}$.
The integral in eq. (\ref{eqn:Eliasheq}) means the surface integral on FS$j$.
The paring interaction $V$ in eq. (\ref{eqn:Eliasheq}) is
\begin{multline}
V^{i j} ( \bm{k} , \bm{k}' , \e_n - \e_m) = \sum_{l_i, \alpha \beta} U_{l_1 \alpha, i}^{*} ( \bm{k} ) U_{l_4 \beta, i} ( \bm{k} ) \\
\times V_{l_1 l_2,l_3 l_4}^{\alpha \beta} (\bm{k} - \bm{k}' , \e_n - \e_m )
U_{l_2 \alpha, j} ( \bm{k}' ) U_{l_3 \beta, j}^{*} ( \bm{k}' ),
\end{multline}
\begin{gather}
\hat{V} = \hat{V}^{\mathrm{c}} + \hat{V}^{\mathrm{s}} + \hat{V}^{(0)}
, \label{eqn:Worb} \\
\hat{V}^{\mathrm{c}} = \frac{1}{2} \hat{\Gamma}^{\mathrm{c}} \hat{\chi}^{\mathrm{c}} \hat{\Gamma}^{\mathrm{c}}, \ \ \ 
\hat{V}^{\mathrm{s}} = -\frac{3}{2} \hat{\Gamma}^{\mathrm{s}} \hat{\chi}^{\mathrm{s}} \hat{\Gamma}^{\mathrm{s}} \label{eqn:Worb-cs} \\
\hat{V}^{(0)} = \frac{1}{2} ( \hat{\Gamma}^{\mathrm{c}}-\hat{\Gamma}^{\mathrm{s}})
\end{gather}
where $U_{l \alpha, i} ( \bm{k} ) = \langle \bm{k} ; l \alpha | \bm{k} ; i \rangle$
is the transformation unitary matrix
between the band and the orbital representations.

In eq. (\ref{eqn:Eliasheq}), $Z$ is given as
\begin{equation}
Z^{i} ( \bm{k} , \e_n ) = 1 + \frac{\gamma^{i}( \bm{k},\e_n )}{| \e_n |},
\end{equation}
where $\gamma^i$ is the impurity induced quasiparticle damping rate.
Here, we calculate the damping rate using $T$-matrix approximation.
We consider the case of Fe-site substitution, 
where the impurity potential $I$ is diagonal in the $d$-orbital basis.
The  $T$-matrix for an impurity at $\alpha(=A\ {\rm or}\ B)$ site is given as
\begin{equation}
\hat{T}^{\alpha} ( \e_n ) = \left[ \hat{1} - \hat{I}^{\alpha} \hat{G}^{\alpha}_{\mathrm{loc}} ( \e_n ) \right]^{-1} \hat{I}^{\alpha} ,
\end{equation}
which is $\k$-independent in the orbital basis.
Here, $I^{\alpha}_{l ,l'} = I \delta_{l,l'}$ is the impurity potential,
and $\hat{G}_{\mathrm{loc}}^{\a}$ is the local Green function given as
\begin{multline}
[G_{\mathrm{loc}}]^{\a}_{l l'} ( \e_n ) = \frac{1}{N} \sum_{\bm{k}'} G_{l l'}^{\a} ( \bm{k}' , \e_n ) \\
= -s_n \frac{i \pi}{( 2 \pi)^3} \sum_{j} \int_{\mathrm{FS}j} \frac{d \bm{k}_{\mathrm{FS}j}'}{v^{j} (\bm{k}' )}
U_{l \a, j} ( \bm{k} ') U_{l' \a, j}^{*} ( \bm{k}' )  ,
\end{multline}
where $s_n = \mathrm{sgn} ( \e_n )$.

In the $T$-matrix approximation, which is exact for $n_{\rm imp}\ll1$,
the normal self-energy in the band diagonal basis is given as
\begin{equation}
\delta \Sigma_{\mathrm{n}}^i ( \bm{k}, \e_n ) = n_{\mathrm{imp}} \sum_{l l' \alpha} U_{l \alpha, i}^{*} ( \bm{k} ) T^{\alpha}_{l l'} ( \e_n ) U_{l' \alpha, i} ( \bm{k} ) ,
\end{equation}
where $n_{\rm imp}$ is the impurity concentration ratio.
Then, the quasiparticle damping rate is given as
\begin{equation}
\gamma^{i} (\bm{k},\e_n)=- \mathrm{Im} \delta \Sigma^{i}_{\mathrm{n}} (\bm{k},\e_n)s_n.
\end{equation}
Also, $\delta \Sigma^i_{\mathrm{a}}$ is the impurity-induced anomalous self-energy given as
%
%
\begin{multline}
\delta \Sigma^i_{\mathrm{a}} ( \bm{k} , \e_n ) = n_{\mathrm{imp}} \sum_{l l' \alpha} U_{l \alpha, i}^{*} ( \bm{k} ) U_{l' \alpha, i} ( \bm{k} ) \\
\times \sum_{m m'} T^{\alpha}_{lm} ( \e_n ) X_{mm'}^{\alpha} ( \e_n ) T^{\alpha}_{l' m'} ( - \e_n ) ,
\end{multline}
where
\begin{multline}
X_{mm'}^{\alpha} ( i\e_n ) =
\frac{\pi}{(2 \pi )^3} \sum_{j} \int_{\mathrm{FS}j} \frac{d \bm{k}_{\mathrm{FS}j}'}{v^{j} (\bm{k}' )} \\
\times U_{m \alpha, j} ( \bm{k}' ) U_{m' \alpha, j}^{*} ( \bm{k}' ) \frac{\Delta^{j} ( \bm{k}' , \e_n)}{| \e_n |} .
\end{multline}

In this calculation, we simplify the energy dependence of $\hat{V}$.
We assume that $\hat{V}^{\xi}$ ($\xi=\mathrm{c},\mathrm{s}$) can be separated into
the momentum and orbital dependent part $\hat{V}^{\xi} (\bm{k} , \w_l = 0)$
and energy dependent part $g_{\xi}(\w_l )$:
\begin{equation}
\hat{V}^{\xi} (\bm{k} , \w_l) = \hat{V}^{\xi}(\bm{k}, \w_{l} = 0) \times g_{\xi}(\w_l).
\end{equation}
We calculated $\hat{V}^{\xi}(\bm{k}, \w_{l} = 0)$ without approximation.
On the other hand, $g_{\xi}(\w_l )$ is determined approximately as
\begin{equation}
g_{\xi}(\w_l)
= \mathrm{Re} \left[
\frac{V^{\xi}_{\mathrm{max}} (\w_l)}
{V^{\xi}_{\mathrm{max}} ( \w_l = 0) } \right] ,
\end{equation}
where $V^{\xi}_{\mathrm{max}} (0)$ is the largest value of
$V^{\xi, \alpha \beta}_{l_1 l_2 , l_3 l_4} ( \bm{k}, \w_l = 0)$
for any $\alpha , \beta , l_i$, and $\bm{k}$.
It is verified that 
this simplification affects the momentum dependence of
the SC gap functions only quantitatively, although 
the obtained $\lambda_{\mathrm{E}}$ is quantitatively underestimated.
Thus, this approximation would be appropriate for the 
present purpose, that is, the analysis of the anisotropy of the SC gap.

\section{Superconducting Gap}
\label{sec:SC}

In this section, we analyze the linearized Eliashberg equation,
eq.(\ref{eqn:Eliasheq}),
using the 3D model of BaFe$_2$(As,P)$_2$ for $n=6.0$.
Hereafter, we use $32 \times 32 \times 16 \bm{k}$ meshes for 
calculating charge and spin susceptibilities.
We assume that $J=J'$ and $U=U'+2J$, and fix the ratio $J/U=1/6$.
In solving the Eliashberg equation,
we used $40 \times 16 \bm{k}$ points for each Fermi surface sheet
and 512 Matsubara frequencies.
In this paper,
we perform the calculation $T=0.005$ and $\omega_{\mathrm{D}}= 0.02$.

\subsection{$s_{\pm}$-wave SC gap mediated by Spin fluctuations}
\label{sec:SC-pm}
\begin{figure}[!htb]
\includegraphics[width=0.9\linewidth]{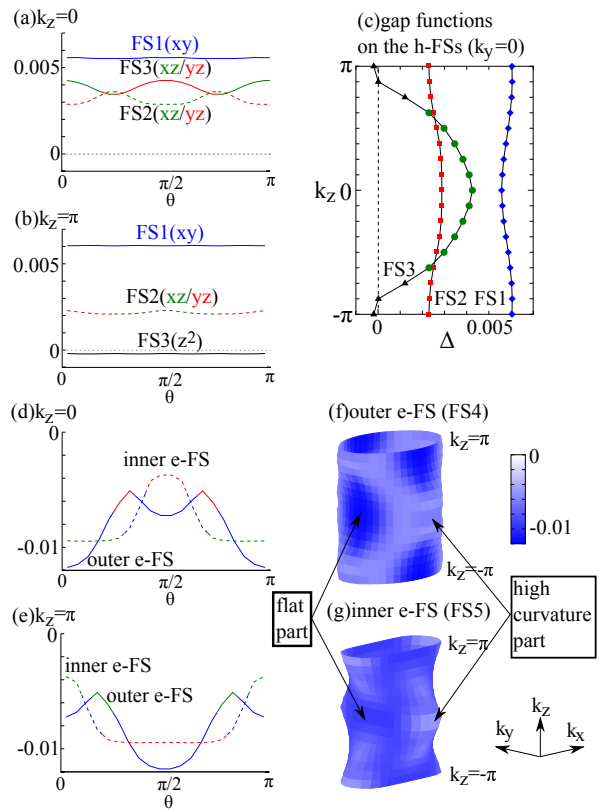}
\caption{
(Color online)
Obtained SC gap functions for $U = 1.15$ and $g = 0$.
(a),(b) SC gap functions on the h-FSs in $k_z=0$ and $k_z= \pi$ planes.
The green, red, blue and black lines correspond to
$xz$, $yz$, $xy$ and $z^2$ orbitals, respectively.
(c) $k_z$ dependence of the SC gaps on the h-FSs in $k_y=0$ planes.
Horizontal node appears on the FS3 around $k_z=\pm\pi$.
(d),(e) SC gap functions on the e-FSs in $k_z=0$ and $k_z= \pi$ planes.
(f),(g) 3D gap functions on the outer and inner e-FSs.
}
\label{fig:s+-gap}
\end{figure}

First, we study the spin-fluctuation-mediated $s_{\pm}$-wave superconducting state
for $U \lesssim U_{\mathrm{cr}}$ by putting $g = 0$ and $n_{\mathrm{imp}} = 0$.
Here, we put $U=1.15$ ($\alpha_{\mathrm{s}} = 0.98$), and the 
obtained eigenvalue is $\lambda_{\mathrm{E}}=1.01$.
The obtained gap structure is almost independent of $\alpha_s$.
First, we discuss the SC gaps on the h-FSs.
Figures \ref{fig:s+-gap}(a) and (b) show the obtained gap functions on the h-FSs
in the $k_z=0$ and $\pi$ planes, respectively.
The definitions of $\theta$ and FS1-5 are shown in Fig. \ref{fig:Fermi}.
In the $k_z=0$ plane, the SC gap size weakly depends on
the orbital character of the FSs.
However, in the $k_z = \pi$ plane, 
the SC gap size strongly depends on the $d$-orbital.
Especially, the SC gap on the $z^2$-orbital FS is almost zero and negative,
reflecting the small spin fluctuations in the $z^2$-orbital
because of the absence of the intra $z^2$-orbital nesting.
(Note that $z^2$-orbital is absent on the e-FSs.)
The horizontal node is clearly recognized in the SC gap
in the $k_y = 0$ plane shown in Fig. \ref{fig:s+-gap} (c).
The obtained  horizontal node on FS3 near $k_z=\pi$ is 
consistent with the previous RPA calculation by Suzuki \textit{et al} \cite{Suzuki}.

The obtained horizontal node  would contradict to 
the four-fold symmetry of the thermal conductivity \cite{Yamashita}
and the small Volovik effect in the specific heat measurement
\cite{Kim,Wang}.
According to ARPES measurements,
the horizontal-node was reported in Ref. \cite{Feng},
whereas it was not observed in Refs. \cite{Shimo-Science,Yoshida}.

Next, we discuss the SC gaps on the e-FSs.
Figures \ref{fig:s+-gap} (d) and (e) show the obtained gap functions 
on the e-FSs in the $k_z=0$ and $\pi$ planes, respectively.
As we can see, line nodes do not appear on the e-FSs.
This result is consistent with the analysis in Ref. \cite{Kuroki-PRB},
that is, the $s_{\pm}$-wave gap on the e-FSs is fully-gapped
if the h-FS made of $xy$-orbital appears.
Note that the SC gaps for $k_z= \pi$ in Fig. \ref{fig:s+-gap} (e) are
obtained by rotating the gaps in the $k_z= 0$ plane in (d)
by $\pi /2$.
Also, Fig. \ref{fig:s+-gap} (f) and (g) show 3D gap functions on the 
outer and inner FSs (FS4 and FS5), respectively.
On both e-FSs, the SC gap on the ``flat part'' is larger
than that on the ``high curvature part''.

\subsection{$s_{++}$-wave SC gap mediated by orbital fluctuations}
\label{sec:SC-pp}

\begin{figure}[!htb]
\includegraphics[width=0.9\linewidth]{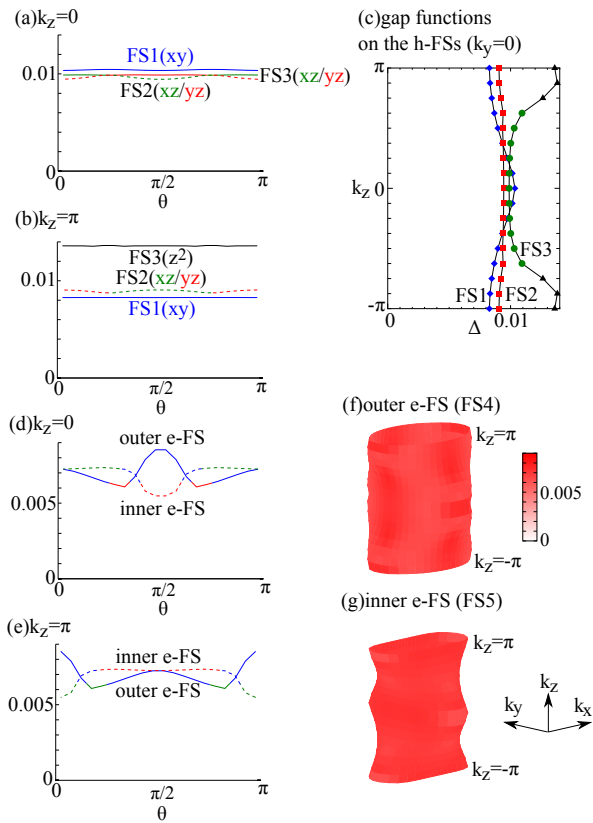}
\caption{
(Color online)
Obtained SC gap functions for $g = 0.22$ and $U = 0$.
(a),(b) SC gap functions on the h-FSs in $k_z=0$ and $k_z= \pi$ planes.
The green, red, blue and black lines correspond to
$xz$, $yz$, $xy$ and $z^2$ orbitals, respectively.
(c) $k_z$ dependence of the SC gaps on the h-FSs in $k_y=0$ plane.
Used colors are same as (a) and (b).
(d),(e) SC gap functions on the e-FSs in $k_z=0$ and $k_z= \pi$ planes.
(f),(g) 3D gap functions on the outer and inner e-FSs.
}
\label{fig:s++gap}
\end{figure}

Next, we study the orbital-fluctuation-mediated $s_{++}$-state superconducting state
for $g \lesssim g_{\mathrm{cr}}$ by putting $U=0$ and $n_{\mathrm{imp}} = 0$.
Here, we put $g=0.22$ ($\alpha_{\mathrm{c}} = 0.98$), and the 
obtained eigenvalue is $\lambda_{\mathrm{E}}=0.59$.
The obtained gap structure is almost independent of $\alpha_c$.
Figures \ref{fig:s++gap} (a) and (b) show the obtained gaps on the h-FSs
in the $k_z=0$ and $\pi$ planes, respectively.
In highly contrast to the spin fluctuation scenario,
the gap size on $z^2$-orbital FS
is comparable with that on the other FSs, since 
strong orbital correlations are developed in all $d$-orbitals:
Note that the quadrupole interaction possesses
many non-zero interorbital matrix elements.
The present numerical result is consistent with our previous calculation
using the 2D 5-orbital model.\cite{Saito-RPA}

Figure \ref{fig:s++gap} (c) shows that
the SC gap size of each h-FSs is approximately independent on $k_z$,
which is consistent with the small orbital dependence of the SC gap
in (Ba,K)Fe$_2$As$_2$ and BaFe$_2$(As,P)$_2$ 
observed in Refs. \cite{Shimo-Science,Yoshida}.
Figures \ref{fig:s++gap}(d) and (e) show the obtained gaps on the e-FSs
in the $k_z=0$ and $\pi$ planes, respectively.
Figure \ref{fig:s++gap} (f) and (g) show the 3D SC gap functions on the 
outer and inner e-FSs (FS4 and FS5), respectively.
Thus, the obtained SC gaps on the e-FSs is isotropic for any $k_z$.

\begin{figure}[!htb]
\includegraphics[width=0.35\linewidth]{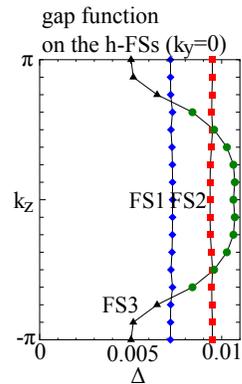}
\caption{
(Color online)
Obtained $k_z$ dependence of the SC gaps on the h-FSs in $k_y=0$ plane.
Used parameters are $g_1=0.24$, $g_2=0$ and $U=0$.
}
\label{fig:noOxy}
\end{figure}

We also discuss the SC gap functions in the case of 
$g_1=g$ and $g_2=0$ in eq. (\ref{eqn:Hint}).
Figure \ref{fig:noOxy} shows $k_z$ dependence of the SC gaps on the h-FSs
for $g=0.24$ ($\alpha_{\mathrm{c}}=0.98$) and $U=0$.
In this case, the gap function on the $z^2$-orbital h-FS is smaller 
compared to the case of $g_1=g_2=g$ in Fig. \ref{fig:s++gap}.
On the other hand,
the obtained SC gaps on the e-FSs are almost isotropic,
similarly to the results of $g_1=g_2=g$.

\subsection{Loop-shape node due to the 
competition of spin and orbital fluctuations}
\label{sec:Loop-shape-node}
Recently, several measurements observed the nodal gap structure
in BaFe$_2$(As$_{1-x}$P$_x$)$_2$ \cite{Yamashita,Shimo-Science,Yoshida}.
This compound is very clean, and very accurate measurements of
gap structure have been performed.
They present a significant challenge for theories
to reproduce the observed gap structure.
However, as discussed in subsection III-A and B,
we cannot reproduce the line-nodes on the electron FSs
when either spin or orbital fluctuations solely develop.

\begin{figure}[!htb]
\includegraphics[width=0.8\linewidth]{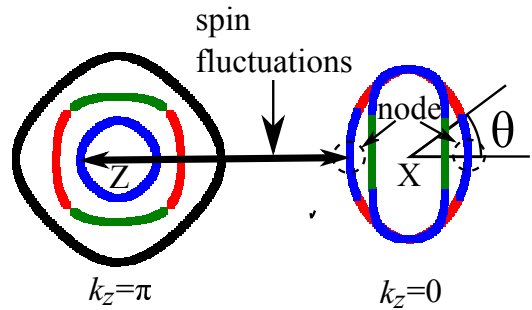}
\caption{
(Color online)
Formation of the nodal $s$-wave gap 
(shown in Fig. \ref{fig:nodalgap1})
due to the competition of orbital fluctuations (=inter-orbital attraction)
and spin fluctuations (=intra-orbital repulsion).
Green, red, blue lines correspond to $xz$, $yz$, $xy$, and $z^2$-orbitals,
respectively.
}
\label{fig:node}
\end{figure}

Here, we study the emergence of highly anisotropic $s$-wave state
due to the strong orbital and spin fluctuations.
In the phase diagram of BaFe$_2$(As$_{1-x}$P$_x$)$_2$,
both $T_N$ and $T_S$ decrease to zero at almost the same critical point
$x_c\approx 0.3$.
This fact means that both spin and orbital fluctuations 
become comparable in magnitude at $x\sim x_c$.
Here, we consider the case that the $s_{++}$-wave state is realized 
by stronger orbital fluctuations.
As increasing the spin fluctuation with momentum $\bm{Q}$,
$\Delta_{\bm{k}}$ and $\Delta_{\bm{k}+\bm{Q}}$ are suppressed
when both $\bm{k}$ and $\bm{k}+\bm{Q}$ are on the FSs with the same orbital character,
and finally the sign change $\Delta_{\bm{k}} \cdot \Delta_{\bm{k}+\bm{Q}}<0$ 
could be achieved.
Such strong anisotropy originates from the competition 
between the attractive interaction of $V^c$ 
and repulsive interaction of $V^s$ in eq. (\ref{eqn:Worb-cs}).
As shown in Fig. \ref{fig:node}, strong spin fluctuations
on the $xy$-orbital (due to intra $xy$-orbital nesting)
produce the loop-shape node on the e-FS.
Similar ''anisotropic $s$-wave gap modified by the spin fluctuations''
is considered to be realized in (Y,Lu)Ni$_2$B$_2$C.\cite{Kontani-B}

\begin{figure}[!htb]
\includegraphics[width=0.9\linewidth]{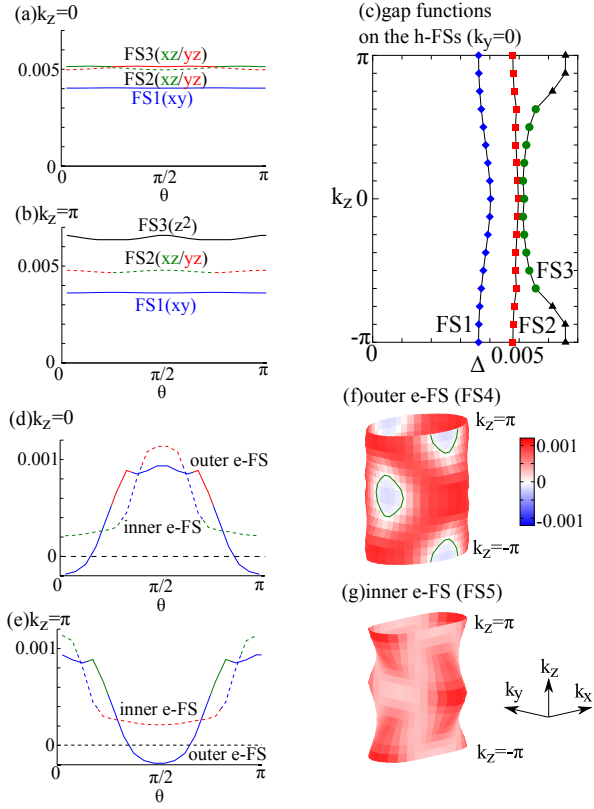}
\caption{
(Color online)
Obtained SC gap functions for $g = 0.204$, $U = 1.011$ and $n_{\mathrm{imp}} = 0.03$.
(a),(b) SC gap functions on the h-FSs in $k_z=0$ and $k_z= \pi$ planes.
(c) $k_z$ dependence of SC gaps on the hole FSs in $k_y=0$ plane.
(d),(e) SC gap functions on the e-FSs in $k_z=0$ and $k_z= \pi$ planes.
(f),(g) 3D gap functions on the outer and inner e-FSs.
The green lines represent the gap nodes.
}
\label{fig:nodalgap1}
\end{figure}

Hereafter, we present numerical results 
in the presence of small amount of impurities ($I$=1 and $n_{\rm imp}=0.03$),
just to make the SC gap functions smoother.
Figure \ref{fig:nodalgap1} show the results of nearly $s_{++}$-wave state
with nodal structure on outer e-FS.
We put $g=0.204$ and $U=1.011$, 
($\alpha_{\mathrm{c}}=0.980$, $\alpha_{\mathrm{s}}=0.859$), and
the eigenvalue is $\lambda_\mathrm{E}=0.50$.
Figures \ref{fig:nodalgap1} (a)-(c) show the obtained SC gaps on the h-FSs
in the $k_z=0$ plane, $k_z=\pi$ plane, and $k_y = 0$ plane, respectively.
The obtained SC gaps on the h-FSs are nearly isotropic and orbital-independent,
similarly to the results in Fig. \ref{fig:s++gap}.
Especially, the gap size of the $z^2$-orbital h-FS is large
even in the presence of loop-shape nodes on e-FSs.

Figures \ref{fig:nodalgap1} (d) and (e) show the obtained SC gaps
on the e-FSs in the $k_z=0$ and $\pi$ planes, respectively.
The SC gap on the inner e-FS is fully opened, and its sign 
is same as that on the h-FSs.
On the outer e-FS, in contrast, the SC gap shows the sign change
near $\theta=0,\pi$ ($\theta=\pi/2, 3\pi/2$) in the $k_z=0$ plane ($k_z=\pi$ plane).
This sign change is caused by strong spin fluctuations in the 
$xy$-orbital, as we have explained in Fig. \ref{fig:node}.
In this case, the SC gaps on the h-FSs remains fully-gapped,
due to the fact that the band-mass of h-FSs is larger than that of the e-FSs.
As results, closed loop-shape nodes appear in the 
flat part on the outer e-FS, as recognized in 
Figs. \ref{fig:nodalgap1} (f) and (g).
This gap structure is consistent with the prediction given by the
 angle-resolved thermal conductivity under the magnetic field
 \cite{Yamashita}.

\begin{figure}[!htb]
\includegraphics[width=0.9\linewidth]{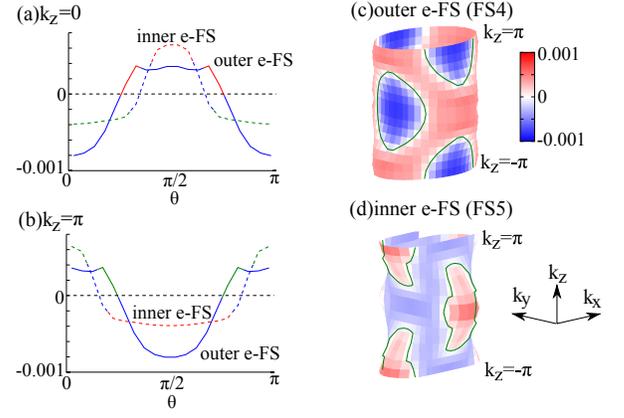}
\caption{
(Color online)
Obtained SC gap functions for $g = 0.204$ and $U = 1.017$.
(a),(b) SC gap functions on the e-FSs in $k_z=0$ and $k_z= \pi$ planes.
(c),(d) 3D gap functions on the outer and inner e-FSs.
The green lines represent the gap nodes.
}
\label{fig:nodalgap3}
\end{figure}

As increasing the value of $U$ (or reducing $n_{\rm imp}$) slightly,
the area of the sign reversed part on the outer e-FS increases, and 
the SC gap on the inner e-FS also shows the sign reversal.
Figures \ref{fig:nodalgap3} (a) and (b) show the SC gap functions on the e-FSs
for $g=0.204$ and $U=1.017$ ($\alpha_{\mathrm{c}}=0.980$ and $\alpha_{\mathrm{s}}=0.864$).
The obtained eigenvalue is $\lambda_{\mathrm{E}} = 0.50$.
The obtained gap functions are approximately given by shifting the gaps
in Figs. \ref{fig:nodalgap1} (d) and (e) downwards,
and line nodes appear on both the inner and outer e-FSs.
As described in Figs. \ref{fig:nodalgap3} (c) and (d),
closed nodal loops appear in the flat part on the outer e-FS
and in the high curvature part on the inner e-FS.
The SC gaps on the h-FSs are almost the same
as those shown in Fig \ref{fig:nodalgap1} (a)-(c),
so we do not show them.

\begin{figure}[!htb]
\includegraphics[width=0.9\linewidth]{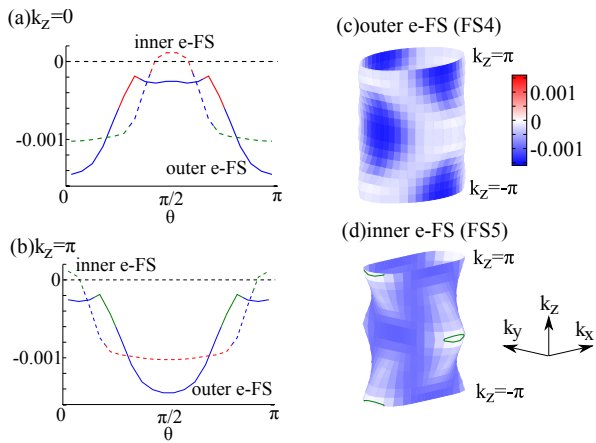}
\caption{
(Color online)
Obtained SC gap functions for $g = 0.204$ and $U = 1.023$.
(a),(b) SC gap functions in the e-FSs on $k_z=0$ and $k_z= \pi$ planes.
(c),(d) 3D gap functions on the outer and inner e-FSs.
The green lines represent the gap nodes.
}
\label{fig:nodalgap2}
\end{figure}

As increasing the value of $U$ (or reducing $n_{\rm imp}$) further,
the sign of the SC gap on the outer e-FS is completely reversed, 
and small closed loop-nodes appear only on the inner e-FS.
The obtained SC gaps are nearly $s_\pm$-wave state.
Figures \ref{fig:nodalgap2} (a) and (b) show the obtained gap functions 
on the e-FSs for $g=0.204$ and $U=1.023$ 
($\alpha_{\mathrm{c}}=0.980$, $\alpha_{\mathrm{s}}=0.869$).
They are approximately given by shifting the gaps
in Figs. \ref{fig:nodalgap3} (a) and (b) downwards.
The obtained eigenvalue is $\lambda_{\mathrm{E}} = 0.50$.
Figures \ref{fig:nodalgap2} (c) and (d) show the obtained 3D gap functions
on the outer and inner e-FSs, respectively.
Apparently, closed nodal loops appear in the high curvature part on the inner e-FS,
whereas no nodes appear on the outer e-FS.
This numerical result is consistent with the recent ARPES measurement 
by Yoshida {\it et al} \cite{Yoshida}.
On the other hand, the SC gaps on the h-FSs are similar to
those in Figs. \ref{fig:nodalgap1} (a)-(c).

\begin{figure}[!htb]
\includegraphics[width=0.80\linewidth]{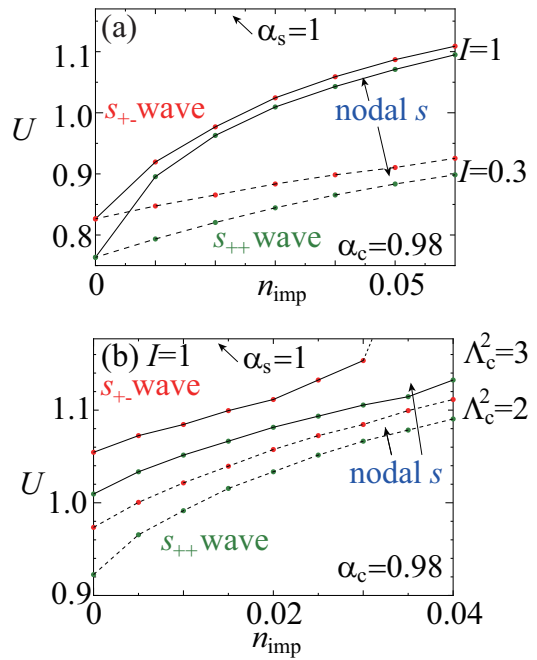}
\caption{
(Color online)
$U$-$n_{\mathrm{imp}}$ phase diagram for $\alpha_{\mathrm{c}}=0.98$
obtained for (a) $I=1$ and $0.3$ with $\Lambda_c=1$
and for (b) $I=1$ with $\Lambda_c=\sqrt{2}$ and $\sqrt{3}$.
}
\label{fig:Unimp}
\end{figure}

In Figs. \ref{fig:nodalgap1}-\ref{fig:nodalgap2},
we fixed the impurity parameters as $n_{\rm imp}=0.03$ and $I=1$.
Now, we discuss the SC gap functions for general impurity parameters.
Figure \ref{fig:Unimp} (a) shows the $U$-$n_{\mathrm{imp}}$ phase diagram 
for both $I=1$ and $I=0.3$.
The solid (dashed) lines represent the boundaries between
$s_{++}$ wave and nodal-$s$ wave, or nodal-$s$ wave and $s_{\pm}$ wave
for $I=1$ ($I=0.3$).
As decreasing $\alpha_{\mathrm{s}}$ or increasing $\alpha_{\mathrm{c}}$,
the following crossover would be realized:
(i) full gap $s_{\pm}$-wave $\to$ (ii) nodal $s$-wave $\to$
(iii) full gap $s_{++}$-wave.
When both $U$ and $g$ are fixed,
the same crossover occurs when $n_{\mathrm{imp}}$ increases.
The residual resistivity for $I=1$ 
derived from the linear response theory 
is about $20 \ \mu\Omega{\rm cm}$ per $n_{\rm imp}=0.01$.

We note that, in the present numerical calculation using 3D model,
line-nodes can appear even if $n_{\mathrm{imp}} = 0$ as shown in 
Fig. \ref{fig:Unimp}.
In contrast, in the previous calculation using 2D model 
\cite{Saito-RPA,Kontani-SSC},
we could not obtain the line-nodes for $n_{\mathrm{imp}} = 0$,
since the SC state changes from the $s_{++}$-wave to $s_\pm$-wave 
discontinuously as $U$ increases.

In the present study based on the RPA,
the $s_{++} \leftrightarrow s_\pm$ crossover is realized
in case of $\a_{\rm s}\ll \a_{\rm c}\ (=0.98)$ for $n_{\rm imp}\ll 0.1$.
One of the  main reasons would be the 
factor $3$ in front of $V^{s}$ in eq. (\ref{eqn:Worb-cs}),
reflecting the SU(2) symmetry of the spin space.
However, this factor $3$ might be overestimated since
the recent polarized neutron scattering measurements indicates the 
relation $\chi^s_z({\bm Q})\gg\chi^s_{x,y}({\bm Q})$ above $T_{\rm c}$
due to the spin-orbit interaction $\lambda {\bm l}\cdot{\bm s}$
\cite{polarized-neutron1,polarized-neutron2}.

Moreover, we have recently
improve the RPA by including the VC, and found that 
orbital fluctuations strongly develop in the Hubbard model
\cite{Onari-SCVC}.
Then, the orbital susceptibility is 
${\hat \chi}^c(q)=({\hat X}^c(q)+{\hat \chi}^0(q))
(1-{\hat \Gamma}^c({\hat X}^c(q)+{\hat \chi}^0(q))^{-1}$,
where ${\hat X}^c(q)$ is the charge VC for the irreducible susceptibility.
According to Ref. \cite{Onari-SCVC},
the magnitude of the three-point vertex is estimated as 
$\Lambda_c=1+X^c(q)/\chi^0(q) \sim2$, and then
eq. (\ref{eqn:Worb-cs}) would be replaced with
$\hat{V}^{\mathrm{c}} = \frac{1}{2} \Lambda_c^2\hat{\Gamma}^{\mathrm{c}} \hat{\chi}^{\mathrm{c}} \hat{\Gamma}^{\mathrm{c}}$.
Figure \ref{fig:Unimp} (b) shows the $U$-$n_{\mathrm{imp}}$ phase diagram 
for $\Lambda_c=\sqrt{2}$ and $\sqrt{3}$, in case of $I=1$.
We find that the $s_{++}$-wave region is widely extended,
and the nodal-$s$-wave region is also widen.
The obtained gap structure in the crossover regime for $\Lambda_c>1$ 
is the loop-shape nodes shown in 
Figs. \ref{fig:nodalgap1}-\ref{fig:nodalgap2}.

\section{discussions}
\label{sec:discussion}

In previous sections, we analyzed the 
gap equations based on the three-dimensional 
ten orbital model for BaFe$_2$(As,P)$_2$.
When orbital fluctuations are solely developed,
fully-gapped $s_{++}$-wave state is realized.
On the other hand, when spin fluctuations are solely developed,
we obtain the $s_\pm$-wave state with horizontal node on a h-FS.
During the crossover between $s_{++}$-wave and $s_{\pm}$-wave states
due to the competition between orbital and spin fluctuations, 
the loop-shape nodes appear on the outer (inner) e-FS 
when the spin fluctuations are slightly weaker (stronger)
than the orbital fluctuations.
The obtained phase-diagram is shown in Fig. \ref{fig:Unimp}.
We stress that all three h-FSs are fully-gapped during the crossover,
since the SC gap on the $z^2$-orbital originates from the 
inter-orbital nesting between different h-FSs.

The crossover from the 
$s_\pm$-wave state to the $s_{++}$-wave state
is also induced by increasing the impurity concentration.
In this study, we considered the 
orbital-diagonal on-site impurity potential at Fe $i$-site,
considering the Fe-site substitution by other elements.
In this case, inter-band impurity scattering
is always comparable to intra-band one, 
as shown by the $T$-matrix approximation in Ref. \cite{Onari-imp}.
For this reason, when the spin fluctuations are solely developed,
the realized $s_{\pm}$-wave state with $T_{\rm c0}\sim30$K
is suppressed by small amount of impurities,
with small residual resistivity $\rho_0\sim 5z^{-1} \mu\Omega$cm
($z^{-1}=m^*/m$ is the mass-enhancement factor).
Since $z^{-1}\sim3$, we can safely expect that
the SC state in {\it dirty Fe-based superconductors}
(say $\rho_0\sim 100\mu\Omega$cm) 
would be the $s_{++}$-wave state due to orbital fluctuations.

There are many important future issues.
As we discussed in Sec. \ref{sec:Loop-shape-node},
one of our important future problems is to study the present 3D ten-orbital 
Hubbard model ($g=0$) using the SC-VC method developed in 
Ref.  \cite{Onari-SCVC}.
Using this method, we have recently shown that the $s_{++}$-wave state 
is realized in the 2D five-orbital Hubbard model ($g=0$).
Note that the ferro-orbital fluctuations induced by the VC,
which explain the orthorhombic structure transition in 
Fig. \ref{fig:phasediagram}, enlarge $T_{\rm c}$ further
\cite{Onari-SCVC2}.
It was recently confirmed that the mechanism of orbital fluctuations due to VC 
had been realized even in a simple two-orbital model,
using the SC-VC method as well as the recently developed 
two-dimensional renormalization group analysis (RG+cRPA method)
\cite{Tsuchiizu,Ohno-SCVC}.

Another important future issue is to include the self-energy 
due to orbital and spin fluctuations, $\Sigma^s$ and $\Sigma^c$, 
into the gap equation.
They are given as
\begin{eqnarray}
{\hat \Sigma}^\xi(k)= T\sum_p (\pm)\hat{V}^\xi(p){\hat G}(k+p),
\end{eqnarray}
where positive (negative) sign corresponds to $\xi=c$ ($\xi=s$).
The real and imaginary parts of the total self-energy
${\hat \Sigma}(k)={\hat \Sigma}^c(k)+{\hat \Sigma}^s(k)$
represent the mass enhancement and quasiparticle inelastic scattering, respectively.
Both effects suppress $T_{\rm c}$.
Moreover, orbital and momentum dependence of ${\hat \Sigma}(k)$
would strongly modify the anisotropy of the SC gap functions.
Thus, the self-energy correction in the gap equation
will be important for the quantitative analysis of the SC gaps.

\section{summary}
\label{sec:summary}

In this paper, we studied SC gap structure using ten orbital model for
BaFe$_2$(As,P)$_2$.
When the orbital fluctuations due to inter-orbital
quadrupole interaction (\ref{eqn:Hint}) are strong, 
the $s_{++}$-wave state is realized.
In contrast, the $s_\pm$-wave state is formed by strong spin fluctuations,
mainly due to intra-orbital Coulomb interaction $U$.
Both spin and orbital fluctuations would strongly develop
in the optimally-doped regime near the O phase.
In this case, we find that a smooth crossover between 
$s_{++}$- and $s_\pm$-wave states is realized by changing the interactions
or impurity concentration, without large suppression in $T_{\rm c}$.

During this $s_{++}\leftrightarrow s_\pm$ crossover,
the loop-shape nodes are universally formed on the e-FSs,
as a result of the competition between inter-orbital attractive interaction
and intra-orbital repulsive interaction.
This result is consistent with recent angle-resolved
thermal conductivity measurement \cite{Yamashita}
and ARPES measurement \cite{Yoshida}.
During the crossover, the SC gaps on the h-FSs are fully-gapped and almost 
orbital independent due to orbital fluctuations, consistently with
recent ARPES measurements \cite{Shimo-Science,Yoshida}.

\acknowledgments
We are grateful to Y. Matsuda, T. Shibauchi, A. Fujimori, T. Yoshida, 
S. Shin, T. Shimojima, D. S. Hirashima and Y. Yamakawa
for valuable discussions. 
This study has been supported by Grants-in-Aid for Scientific 
Research from MEXT of Japan.
Numerical calculations were partially performed using 
the Yukawa Institute Computer Facility.

\appendix
\section{orabital susceptibilities in the present model}
\label{sec:orbital-fluctuations}

In this paper, we discussed the development of orbital fluctuations
due to the quadrupole interaction in eq. (\ref{eqn:Hint}).
Here, we consider the quadrupole operator at $i=(a,\alpha)$,
where $a$ and $\alpha$ represent the unit cell and the 
Fe-site (A or B), respectively.
Then, the operator ${\hat O}^i_{\Gamma}$ ($\Gamma=xz,yz,xy$) is given as
\begin{eqnarray}
{\hat O}^i_{\Gamma}\equiv \sum_{lm} o^{l,m}_\Gamma {\hat m}_{l,m}^i ,
\end{eqnarray}
where $l$ and $m$ represents the $d$-orbital,
${\hat m}_{l,m}^i \equiv \sum_\s c_{li\s}^\dagger c_{mi\s}$,
and the coefficient is defined as
$o^{l,m}_{xz}= 7\langle l|{\hat x}{\hat z}|m \rangle$
for $\Gamma=xz$, where ${\hat x}=x/r$ and so on.
The non-zero coefficients are given as \cite{Kontani-AL}
\begin{eqnarray}
&& o^{2,5}_{xz}=o^{3,4}_{xz}=\sqrt{3}o^{1,2}_{xz}=1,
 \label{eqn:oxz}\\
&& -o^{3,5}_{yz}=o^{2,4}_{yz}=\sqrt{3}o^{1,3}_{yz}=1,
  \label{eqn:oyz}\\
&& o^{2,3}_{xy}=-\sqrt{3}o^{1,4}_{xy}/2=1.
 \label{eqn:oxy}
\end{eqnarray}
where 1,2,3,4,5 respectively correspond to $z^2$, $xz$, $yz$, $xy$, $x^2-y^2$.

\begin{figure}[!htb]
\includegraphics[width=0.9\linewidth]{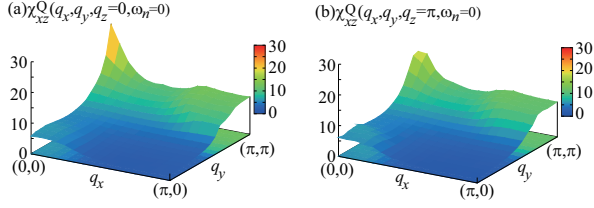}
\caption{
(Color online)
Obtained $O_{xz}$-channel quadrupole fluctuations $\chi_{xz}^{Q}(\bm{q},0)$
for $n=6.0$ and $\alpha_c=0.98$, in the 
(a) $q_z=0$ plane and (b) $q_z=\pi$ plane.
The obtained $q_z$ dependence of $\chi_{xz}^{Q}(\bm{q},0)$ is rather weak.
}
\label{fig:chiQ}
\end{figure}

In the presence of quadrupole interaction in eq. (\ref{eqn:Hint}),
${\hat \chi}^c(q)$ given in eq. (\ref{eqn:chic}) is strongly enhanced.
Then, it is convenient to introduce the 
quadrupole interaction, by 
${\hat \chi}^{Q,\alpha\beta}_{\Gamma}(q)
\equiv {\hat o}_{\Gamma} {\hat \chi}^{c,\alpha\beta}(q) {\hat o}_{\Gamma}\}$,
where $\alpha,\beta=$A or B.
In the present study, the channel $\Gamma=xz,yz$
is the most strongly enhanced, due to the good inter-orbital 
nesting in Fe-based superconductors \cite{Kontani-RPA}.
Figure \ref{fig:chiQ} shows the obtained 
${\hat \chi}^{Q}_{xz}(\bm{q},0)
={\hat \chi}^{Q,AA}_{xz}(\bm{q},0)+{\hat \chi}^{Q,AB}_{xz}(\bm{q},0)$
in the $q_z=0$ and $q_z=\pi$ planes
\cite{Kontani-AL}.
The large peak at $\bm{q}\approx (0,\pi)$ originates from the 
interorbital ($yz\leftrightarrow xy$) nesting between e-FS and h-FS,
and the small peak at $\bm{q}\approx (\pi,\pi)$ originates from the 
interorbital ($xz\leftrightarrow z^2$) nesting between two h-FSs.
Therefore, the obtained development of
${\hat \chi}^{Q}_{xz}(q)$ and ${\hat \chi}^{Q}_{yz}(q)$ 
means the existence of strong orbital fluctuaitions
on the $xz$-, $yz$-, $xy$- and $z^2$-orbitals, of which 
the FSs of BaFe$_2$(As,P)$_2$ are composed.
For this reason, we obtain the orbital-fluctuaition-mediated
$s_{++}$-wave state with approximately isotropic SC gap
on the three h-FSs, consistently with the ARPES measurements 
in Refs. \cite{Shimo-Science,Yoshida}.

In addition to the AF orbital fluctuation, 
the ferro-orbital fluctuations with respect to $O_{x^2-y^2}= n_{xz}-n_{yx}$ 
are induced by the "two-orbiton process" in Ref \cite{Kontani-AL} 
as well as "two-magnon process" in Ref. \cite{Onari-SCVC}.
These processes are given by the Azlamasov-Larkin (AL) 
type vertex correction, since the AL term describes the interference 
between ferro- and AF-fluctuations that is neglected in the RPA.
The ferro-orbital fluctuations induce the softening of $C_{66}$
as well as the orthorhombic structure transition.
Although the ferro-orbital fluctuations also contribute to
the $s$-wave SC state, we consider that they are not the major 
mechanism of the SC: 
First, the relation $\Delta_{xz}, \Delta_{yz} \gg \Delta_{xy}$ is expected 
in the ferro-orbital fluctuation mediated $_{s++}$ wave state, 
since these fluctuations develop only on the $xz$ and $yz$ orbitals. 
This relation is inconsistent with experiments. 
Second, we have recently solved the Eliashberg gap equation based on 
the SC-VC method \cite{Onari-SCVC2}
and found that the main pairing interaction is actually 
AF orbital fluctuations with respect to $O_{xz,yz}$, 
since the peak of the ferro-orbital susceptibility is very narrow 
in the momentum space. 
Therefore, the present analysis by taking only AF orbital fluctuations 
would be justified.

\section{heavily electron-doped case}
\label{sec:heavily-doped}

\begin{figure}[!htb]
\includegraphics[width=0.9\linewidth]{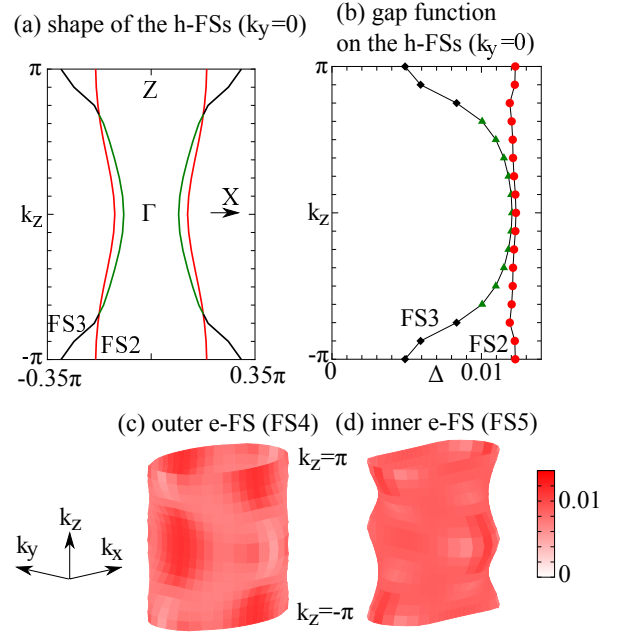}
\caption{
(Color online)
Obtained SC gap functions in the over-doped case ($n = 6.1$)
for $g = 0.25$, $U = 0$ and $n_{\mathrm{imp}} = 0$.
(a) $k_z$-dependence of the h-FSs, and 
(b) the obtained SC gap functions on the h-FSs.
We also show the 3D gap functions on the 
(c) outer e-FS and (d) inner e-FS.
}
\label{fig:gap_n61_s++}
\end{figure}

In this paper, we have studied the gap functions for $n=6.0$.
In this Appendix, we discuss the heavily electron doped case ($n=6.1$).
Figure \ref{fig:gap_n61_s++} (a) shows the h-FSs on the $k_y=0$ plane.
As one can see, the $xy$-orbital h-FS disappears.

Now, we consider the orbital-fluctuation-mediated $s_{++}$-wave SC state.
Here, we put $g = 0.25$ ($\alpha_{\mathrm{c}} = 0.98$) and $U=0$
and obtained eigenvalue $\lambda_{\mathrm{E}}$ is 0.87.
First, we discuss the SC gaps on the h-FSs.
Figure \ref{fig:gap_n61_s++} (b) shows the $k_z$ dependence of the 
SC gaps on the h-FSs in the $k_y=0$ plane.
In contrast to the case of $n=6.0$,
the SC gap on the $z^2$-orbital hole FS becomes smaller, while
each SC gap on the h-FSs on any plane perpendicular to the $k_z$-axis
is almost isotropic.
Next, we focus on the SC gaps on the e-FSs.
Figures \ref{fig:gap_n61_s++} (c) and (d) show
the 3D gap functions on the outer and inner e-FSs, respectively.
As with the case of $n=6.0$,
the SC gaps on the e-FSs are nearly isotropic
on any plane perpendicular to the $k_z$-axis.

\begin{figure}[!htb]
\includegraphics[width=0.9\linewidth]{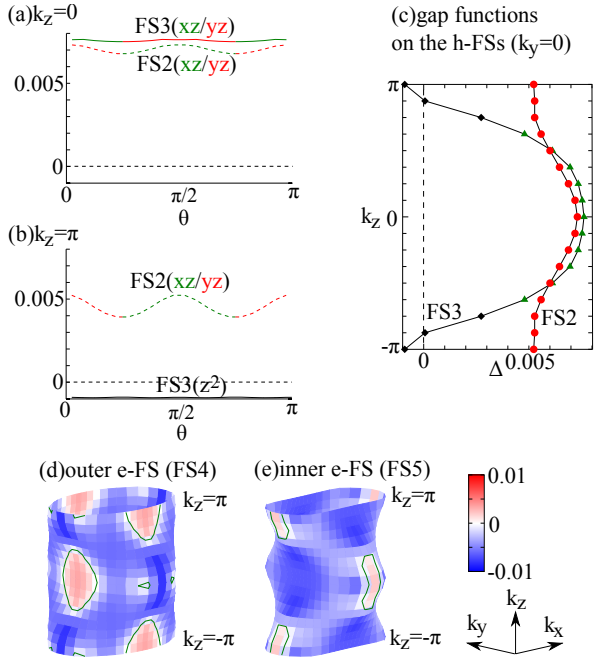}
\caption{
(Color online)
Obtained SC gap functions in the over-doped case ($n=6.1$)
for $g = 0$, $U = 1.37$ and $n_{\mathrm{imp}} = 0$.
(a),(b) SC gap functions on the h-FSs in the $k_z=0$ and $k_z= \pi$ planes.
(c) $k_z$ dependence of SC gap functions on the h-FSs for $k_y=0$.
(d),(e) 3D gap functions on the outer e-FS and inner e-FS.
}
\label{fig:gap_n61_s+-}
\end{figure}

Next, we discuss the spin-fluctuation-mediated $s_{\pm}$-wave SC state.
Here, we put $U = 1.37$ ($\alpha_{\mathrm{s}} = 0.98$) and $g = 0$.
The obtained eigenvalue $\lambda_{\mathrm{E}}$ is 1.68.
Figures \ref{fig:gap_n61_s+-} (a) and (b) show the gap functions on the h-FSs
in the $k_z=0$ and $k_z= \pi$ planes, respectively.
In common with the results of $n=6.0$,
the SC gap function on the $z^2$-orbital h-FS is almost zero
since the spin fluctuations in the $z^2$-orbital do not develop.
Figures \ref{fig:gap_n61_s+-} (c) shows the $k_z$ dependence on the SC gaps 
on the h-FSs in the $k_y=0$ plane.
It clearly shows the horizontal line node on FS3 near $k_z= \pi$.

Next, we discuss on the e-FSs.
Figures \ref{fig:gap_n61_s+-} (d) and (e) show the 3D gap functions
on the outer and inner e-FSs, respectively.
Unlike the case of $n=6.0$, line nodes appear on the e-FSs even when $g=0$.
This result is consistent with the analysis in Ref. \cite{Kuroki-PRB}:
nodal gap appears on the e-FSs
when the $xy$-orbital h-FS disappears
because of the competition of different spin fluctuations;
$\bm{Q} = ( \pi , 0 )$ on $xz+yz$-orbitals 
and $\bm{Q} = (\pi , \pi /2 )$ on $xy$-orbital.
In the presence of $xy$-orbital h-FS for $n \approx 6.0$,
on the other hand, spin fluctuations develop at $\bm{Q} = ( \pi ,0)$ 
in all $d$-orbitals.
Then, fully-gapped $s_{\pm}$-wave state is realized.
In this case, the competition of orbital and spin fluctuations induce
the loop-shape nodes as discussed in Sec. III-C.

We comment that Khodas and Chubukov had discussed the emergence of the 
loop-shape nodes on the e-FSs in the ``folded model'' with
two Fe-atoms in each unit-cell \cite{Khodas}:
When the ``vertical nodes'' of e-pockets are realized in the 
``unfolded model'' with one Fe-atoms in each unit-cell, 
the loop-shape nodes are realized in the folded model 
by taking the finite hybridization between two e-FSs 
inherent in 122 systems \cite{Saito-KFe2Se2} into account.
Within the RPA, the vertical nodes appear only in the absence of 
the xy-like hole pocket at $(\pi,\pi)$ as discussed by 
Kuroki et al. \cite{Kuroki-PRB}, 
and therefore the mechanism of the "loop-shape nodes" by 
Khodas and Chubukov requires the absence of xy-like hole pocket, 
at least within the RPA. 
The calculation in this Appendix gives a numerical verification 
of the theory of Khodas and Chubukov.


\end{document}